\documentstyle[12pt,aaspp4]{article}
\def\Msun{~M_\odot}
\def\lsim{\raise0.3ex\hbox{$<$}\kern-0.75em{\lower0.65ex\hbox{$\sim$}}}
\def\gsim{\raise0.3ex\hbox{$>$}\kern-0.75em{\lower0.65ex\hbox{$\sim$}}}
\def\gcm{\rm ~g~cm^{-3}}
\def\kms{\rm ~km~s^{-1}}

\def\ml{~\Msun ~\rm yr^{-1}}

\begin{document}

\title{GAMMA-RAY BURST ENVIRONMENTS AND PROGENITORS}
\author{Roger A. Chevalier and Zhi-Yun Li}
\affil{Department of Astronomy, University of Virginia, P.O. Box 3818}
\affil{Charlottesville, VA 22903; rac5x@virginia.edu, zl4h@virginia.edu}


\begin{abstract}

Likely progenitors for the  GRBs (gamma-ray bursts) are
the mergers of  compact objects or the explosions of massive stars.
These two cases have distinctive environments for the GRB afterglow:
the compact object explosions occur in the ISM (interstellar medium) and those
of massive stars occur in the preburst stellar wind.
We calculate the expected afterglow for a burst in a Wolf-Rayet
star wind and compare the results to those for constant, interstellar
density.
The optical afterglow for the wind case is generally expected to decline
more steeply than in the constant density case, but this effect may be
masked by variations in electron spectral index, and the two cases have
the same evolution in the cooling regime.
Observations of the concurrent radio and optical/X-ray evolution are
especially useful for
distinguishing between the two cases.
The different rates of decline of the optical and X-ray 
afterglows of GRB 990123 suggest constant density interaction for this case.
We have previously found strong evidence for wind interaction in
SN 1998bw/GRB 980425 and here present a wind model for GRB 980519.  
We thus suggest that there are both wind type GRB afterglows with
massive star progenitors  and ISM
type  afterglows with compact binary star progenitors.
The wind type bursts are likely to be accompanied by a supernova, but
not the ISM type.

\end{abstract}


\section{INTRODUCTION}

Although the study of GRBs (gamma-ray bursts) has been revolutionized
in the past few years by finding a number of precise positions and distances,
 the nature of their progenitor objects remains uncertain
 (see M\'esz\'aros 1999 for a review).
The production of a large amount of energy in a short time has naturally
led to models involving compact objects (neutron stars and black holes).
Fryer, Woosley, \& Hartmann (1999) have summarized possible progenitors
involving black hole accretion disks: neutron star - neutron star binary mergers
(NS/NS), black hole - neutron star mergers (BH/NS), black hole - white
dwarf mergers (BH/WD), massive star core collapses, and
black hole - helium star mergers (BH/He).
Based on estimated formation rates and on accretion disk models with
high viscous forces, Fryer et al. (1999) suggest that NS/NS and BH/NS
mergers dominate the population of short-duration GRBs and that
massive stars and BH/He mergers dominate the long-duration bursts.
The afterglows observed to date would have massive star progenitors in
this scenario because they followed long-duration bursts.

One way of distinguishing between the progenitor models is to examine the
position of the GRB in the parent galaxy.
Paczy\'nski (1998) pointed out that NS/NS binaries would be expected to
have a significant space velocity, which would carry them many kpc from
their birthplaces.
The observational evidence for the association of several GRBs with star
forming regions then provided weak evidence against the NS/NS merger
progenitors and favored massive star progenitors.
The population synthesis calculations of Fryer et al. (1999) supported
this conclusion.
However, Bloom, Sigurdsson, \& Pols (1999a) found 
a time to NS/NS merger of $\sim 10^8$ years.
These objects would then follow the star formation rate, although
$\sim 15$\% of them might occur well outside of dwarf galaxy hosts.

The connection of GRBs to massive stars became stronger with the
discovery of the Type Ic supernova SN 1998bw in the error box of
GRB 980425 (Galama et al. 1998). 
The high energy inferred for the optical supernova,
$(2-3)\times 10^{52}$ ergs (Iwamoto et al. 1998; Woosley, Eastman,
\& Schmidt 1999), and the high expansion velocity inferred for the radio
supernova (Kulkarni et al. 1998) strengthen the GRB connection.
Li \& Chevalier (1999) found that the evolution of the radio source indicated
non-uniform energy input to the blast wave, as is also inferred in GRBs.
They also found evidence that the radio SN 1998bw
interacted with the stellar wind expected from the massive star progenitor.
In this paper, we emphasize that a stellar wind environment is an
unavoidable consequence of a massive star progenitor and that the nature
of the GRB afterglow emission can provide a discriminant between massive
star and compact binary progenitor models.
In \S~2, we discuss the expected afterglow for a massive star progenitor
model and in \S~3 place our models in the context of observations.
Our discussion concentrates on the cases $s=2$ (stellar wind) and
$s=0$ (interstellar medium), where the ambient medium has $\rho\propto
r^{-s}$.

\section{AFTERGLOWS IN A MASSIVE STAR WIND}

In the existing massive star GRB progenitor models, the most likely
progenitor is the stripped core of a massive star, i.e., a Wolf-Rayet star.
MacFadyen \& Woosley (1999) consider single massive stars whose cores
directly collapse to black holes.
The stars have an initial 
mass $\gsim 25\Msun$, which is the type of star that is
likely to lose its H envelope in winds.
Paczy\'nski (1998) noted that the requirement of a rapidly rotating
core might necessitate a close binary companion, which again points to
 Wolf-Rayet stars.
The winds from these stars in our Galaxy have velocities of $1,000-2,500\kms$
and mass loss rates $\dot M\approx 10^{-5}-10^{-4}\ml$ (Willis 1991).
On evolutionary grounds, Langer (1989) advocated 
 $\dot M \sim 6\times 10^{-8}(M_{\rm WR}/\Msun)^{2.5}\ml$, where $M_{\rm WR}$ is
the mass of the Wolf-Rayet star.
If the stellar mass drops to $\sim 3\Msun$ at the end of its life because of
mass loss,  $\dot M\sim 10^{-6}\ml$ at that time.
The host galaxies of GRBs are likely to be of low metallicity.
Willis (1991) notes that there is no evidence for a metallicity dependence
of mass loss from a particular type of Wolf-Rayet star, but that metallicity
may affect the distribution of Wolf-Rayet types.

The effects of Wolf-Rayet winds can be observed in the wind bubbles around some
of these objects.
In some cases where the bubble is well-studied, the bubble expansion 
and X-ray emission
are consistent with a weaker wind than that inferred from direct
observations of the Wolf-Rayet star (e.g., NGC 6888, 
Garc\'ia-Segura \& Mac Low 1995).
The position of the wind
termination shock, $r_t$, can be estimated by equating
 the ram pressure in the wind with the pressure in the  bubble.
We estimate $r_t/R_b\approx (2R_b/v_wt_w)^{1/2}$, where $R_b$ is the radius of
the wind bubble and $t_w$ is its age.
For $R_b=3$ pc, $t_w=2\times 10^4$ yr, and $v_w=1,000\kms$, we have
$r_t\approx 5\times 10^{18}$ cm.
At $r_t$, the density  increases by a factor of 4 and becomes
approximately constant at larger radius.
The radius $r_t$ is expected to increase as the bubble evolves and its
pressure drops.
At the end of its life, a Wolf-Rayet star is thus expected to be surrounded
by a substantial $\rho\propto r^{-2}$ medium.
The radio observations of Type Ib/c supernovae can be interpreted as 
interactions with such a medium in models with constant efficiencies of 
production of relativistic electrons and magnetic fields
(Weiler et al. 1999; Chevalier 1998).

For the wind density, we take $\rho
=Ar^{-2}$, where $A=\dot M/4\pi v_w=5\times 10^{11}A_{\star}$ g cm$^{-1}$.
The reference value of $A$ corresponds to $\dot M=1\times 10^{-5}\ml$
and $v_w=1000\kms$.
If a burst were able to occur in a red supergiant star,  the low
wind velocity, $v_w\approx 10\kms$, would lead to a higher circumstellar
density because $\dot M\gsim 10^{-7}\ml$ up to $\sim 10^{-4}\ml$
is expected.
The only supernova case where we have inferred a lower circumstellar
density is around SN 1987A, which exploded as a B3 I star and was thus
too cool to drive a strong wind with its ultraviolet radiation field.
In this case, the $r^{-2}$ wind terminated at $\sim (3-4)\times 10^{17}$ cm
and the supernova radio flux rose sharply when the shock front encountered
denser gas
(Ball et al. 1995; Chevalier  1998).

The scaling laws that are appropriate for GRB interaction with an $s=2$ 
medium have been described by M\'esz\'aros, Rees, \& Wijers (1998) and by
Panaitescu, M\'esz\'aros, \& Rees (1998).
Our aim here is to examine the specific predictions for interaction
with a Wolf-Rayet star wind.
We calculate the expected emission for a thin shell model (cf. Li \&
Chevalier 1999).
We adopt a nucleon-to-electron number density ratio of two, which 
is appropriate for winds of Wolf-Rayet stars that are predominantly 
helium (and perhaps carbon/oxygen). 
For the mass loss case, a density of $1.67\times 10^{-24}\gcm$ is attained at
$r\approx 5.5\times 10^{17}A_{\star}^{1/2}$ cm.
For an adiabatic blast wave in an $s=2$ medium (Blandford \& McKee 1976), 
we find $\gamma^2 =R/4ct$ and $R=(9Et/4\pi Ac)^{1/2}=
 2.0\times 10^{17} E_{52}^{1/2}
A_{\star}^{-1/2}t_{\rm day}^{1/2}$ cm, where $\gamma$ is the Lorentz factor
of the gas, $R$ is the observed radius near the line of sight,
$E_{52}=E/10^{52}$ ergs is the explosion energy,
and $t=t_{\rm day} {\rm ~day}$  is the time in the observer's
frame.
Over the typical time of observation of a GRB afterglow, 
the shock front is expected
to be within the $s=2$ wind, except for the unusual case of a progenitor
like that of SN 1987A.

In our model, the synchrotron emission frequency of the lowest energy electrons
is
\begin{equation}
\nu_m= 5\times 10^{12}\left(1+z\over 2\right)^{1/2}
(\epsilon_e/0.1)^2(\epsilon_B/0.1)^{1/2}E_{52}^{1/2}
t_{\rm day}^{-3/2} {\rm~Hz},
\end{equation}
and the flux at this frequency is
\begin{equation}
F_{\nu_m}= 20\left(\sqrt{1+z}-1\over \sqrt{2}-1\right)^{-2}
\left(1+z\over 2\right)^{1/2}
(\epsilon_B/0.1)^{1/2} E_{52}^{1/2} A_{\star}
t_{\rm day}^{-1/2} {\rm~mJy},
\end{equation}
where $z$ is the redshift, and $\epsilon_e$ and $\epsilon_B$ are the
electron and magnetic postshock energy fractions.
These expressions assume a flat universe with $H_o=65$ km s$^{-1}$ Mpc$^{-1}$.
As in discussions of afterglows in the ISM, the magnetic field is assumed
to be amplified by processes in the shocked region and is not directly
determined by the field in the Wolf-Rayet star wind.
For electrons with a power law energy distribution above $\gamma_{em}$,
$dN_e/d\gamma_e\propto \gamma_e^{-p}$, the flux above $\nu_m$ is
$F_{\nu}=F_{\nu_m}(\nu/\nu_m)^{-(p-1)/2}
\propto \epsilon_e^{p-1}\epsilon_B^{(p+1)/4}
E^{(p+1)/4} A t^{-(3p-1)/4} \nu^{-(p-1)/2}$,
provided the electrons are not radiative.
$F_{\nu}=F_{\nu_m}(\nu/\nu_m)^{1/3}$ below $\nu_m$ 
until the spectrum turns over due to synchrotron 
self-absorption at
\begin{equation}
\nu_A\approx 1\times 10^{11} \left(1+z\over 2\right)^{-2/5}
(\epsilon_e/0.1)^{-1}(\epsilon_B/0.1)^{1/5}E_{52}^{-2/5}A_{\star}^{6/5}
t_{\rm day}^{-3/5} {\rm~Hz}.
\end{equation}
We estimate the effects of synchrotron cooling following 
the discussions of Sari, Piran, \& Narayan  (1998)
and Wijers \& Galama (1999) for the $s=0$ case.
Radiative cooling becomes important at a frequency
\begin{equation}
\nu_c\approx 1\times 10^{12}\left(1+z\over 2\right)^{-3/2}
(\epsilon_B/0.1)^{-3/2}E_{52}^{1/2}A_{\star}^{-2}
t_{\rm day}^{1/2} {\rm~Hz},
\end{equation}
provided $\nu_c>\nu_m$.
For $\nu>\nu_c$, $F_{\nu}\propto \epsilon_e^{p-1}\epsilon_B^{(p-2)/4}
E^{(p+2)/4} t^{-(3p-2)/4} \nu^{-p/2}$.

These results can be compared to similar results for interaction with
a constant density interstellar medium with $n_o\approx 1$ 
cm$^{-3}$ (e.g., Waxman 1997;
Sari et al. 1998).
In that case $F_{\nu_m}$ is constant with time, 
$\nu_m$ is comparable to the value given above,  $\nu_A$
is independent of time, and $\nu_c$ decreases with time.
If $\nu_A$, $\nu_m$, and $\nu_c$ have the same relation to each other in
the $s=0$ and $s=2$ cases, the appearance of the spectrum at one
time is similar for both cases, but the  evolution is different.
At high frequency (optical and X-ray), for $s=0$ the flux evolution
goes from adiabatic ($\propto t^{-(3p-3)/4}$) to cooling 
($\propto t^{-(3p-2)/4}$) while for $s=2$ the flux evolution
goes from cooling ($\propto t^{-(3p-2)/4}$) to adiabatic 
($\propto t^{-(3p-1)/4}$).
While cooling, the two cases have the same spectrum and flux decline.
At low frequency (radio) with $\nu<\nu_m$, the flux evolves as $t^{1/2}$
for $s=0$, but can make a transition from $\propto t$ to constant for $s=2$.

\section{COMPARISON WITH OBSERVATIONS}

We have noted the evidence that SN 1998bw is interacting with a
circumstellar wind (Li \& Chevalier 1999), and here discuss
 the cosmological GRBs.
The more detailed models that have been developed for comparison with
the best observed GRB afterglows have  taken a low, constant density
surrounding medium as the starting point.
For GRB 970508, Wijers \& Galama (1999) and Granot, Piran,
\& Sari (1999) found $n_o=0.030, 5.3$ cm$^{-3}$, $E_{52}=
3.5, 0.53$, $\epsilon_e=0.12, 0.57$, and $\epsilon_B=0.089,0.0082$,
respectively.
The difference between these results reflects, in part, the difficulty
of accurately determining the significant frequencies, such as $\nu_A$.
The low density found for this case is indicative of an interstellar
density.
However, these models primarily depend on the spectrum at one time,
which can be the same for the $s=0$ and $s=2$ cases.
Panaitescu et al. (1998) have presented a successful $s=0$ model for the
evolution of GRB 970508, but a number of deviations from the simple
model are included, so the case is not clear.

With $s=0$, the optically thin, adiabatic flux evolution can be described by
 $F_{\nu}\propto t^{\alpha}\nu^{\beta}$ with
$\alpha=1.5\beta$.
This type of evolution was observed in the afterglow of 
 GRB 970228 (Wijers, Rees, \& M\'esz\'aros 1997).
For $s=2$, the expected power law evolution has $\alpha=(3\beta-1)/2$
for a constant energy blast wave, so a mechanism must be found
to flatten the time evolution if this GRB evolved
in a circumstellar wind.
The effects of beaming and a change to nonrelativistic flow steepen
the time evolution, but continued power input from ejecta can flatten it.
This effect is inferred to occur in radio supernovae (Chevalier 1998).
Following Rees \& M\'esz\'aros (1998) for the $s=0$ case with power
input from ejecta with a mass gradient $M(\gsim \Gamma_f)\propto
\Gamma_f^{-n}$ where $\Gamma_f$ is the Lorentz factor in the freely
expanding ejecta, the evolution follows $F_{\nu}\propto
t^{-[2-\beta (n+6)]/(n+4)}$ for the $s=2$ case.
For $\beta=-0.75$, the property $\alpha=1.5\beta$ is recovered when $n=5.33$.
The optically thin, $\nu>\nu_m$ evolution would then be the same as
in the constant density ($s=0$) case, but other aspects of the evolution
would be different.
In particular, the expansion would be much less decelerated and the
apparent radius of the blast $r\propto t^{(n+2)/(n+4)}\propto
t^{0.79}$ and its Lorentz
factor $\gamma\propto t^{-1/(n+4)}\propto t^{-0.11}$.
At optically thick wavelengths, the flux would increase as $F_{\nu}\propto
t^{1.57}$ as opposed to $F_{\nu}\propto t^{1/2}$ for the constant density
case.
Radio observations could distinguish this kind of evolution, but GRB 970228
was not detected in the radio.

Kulkarni et al. (1999) noted that the afterglow of GRB 990123 was 
consistent with $p=2.44$ in a $s=0$ model, where $p$ is the electron
energy spectral index.
The optical emission declined with $\alpha=-1.10\pm 0.03$, suggesting
adiabatic evolution, while the X-rays declined with 
$\alpha=-1.44\pm 0.07$, suggesting
cooling evolution.
The faster decline at X-ray wavelengths is distinctive of $s=0$ evolution
and is the opposite of expectations for $s=2$.
Most of the well observed optical afterglows also have $\alpha$ in the range
--1.1 to --1.3, which is plausibly modeled by blast wave evolution in
a constant density medium.
However, the possibilities of flat electron spectra or cooling evolution
do not allow $s=2$ models to generally be ruled out.

An afterglow with a steep decline was that of GRB 980519.
Its optical emission followed $F_{\nu}\propto 
t^{-(2.05\pm 0.04)}$, which, with the observed $\beta=-1.05\pm 0.10$ 
for optical and
X-ray data, is consistent with expansion in an $s=2$ medium (Halpern
et al. 1999).
The steep decline might be the result of beaming (Halpern et al. 1999;
Sari, Piran, \& Halpern 1999) instead of interaction with a wind.
One way of distinguishing the wind interaction case is to observe
the optically thick
radio flux rise, $F_{\nu}\propto t$ for a wind as opposed to 
$F_{\nu}\propto t^{1/2}$ for a constant density.
The effect of beaming in the $s=0$ case
 is to decrease the expansion and thus to increase
the difference between the models.
Radio observations of GRB 980519 were made at ages 0.3, 1.1, and 2.8 days
at 8.3 GHz (Frail et al. 1998).
Fig. 1 shows that an $s=2$ model is capable of fitting the radio, as well
as the optical and X-ray, data.

The model light curves in Fig. 1 are obtained using the  
synchrotron self-absorption model of Li \& Chevalier (1999). The 
model takes into account the dynamical evolution of a spherical, 
constant-energy blast wave in a pre-burst $s=2$ stellar wind, 
relativistic effects on radiation, and synchrotron self-absorption. 
The distance to GRB 980519 
 is unknown, so we take a fiducial value of 
$z=1$. Adopting a 
relatively low magnetic energy fraction of $\epsilon_B=10^{-5}$, we 
find that the following combination of parameters fits all available 
data reasonably well: $E_{52}=0.54$, $\epsilon_e=0.62 $, $p=3$, and 
$A_{\star}=4.3 $. Rough scalings of these
parameters for other choices of $\epsilon_B$ are given in Li 
\& Chevalier (1999). 
The value of $p$ is higher than that
normally found in GRB afterglows, but is within the range found
in radio supernovae (Chevalier 1998 and references therein).
The model shown in Fig. 1 does not include synchrotron cooling.
The good fit to the X-ray data is then expected 
because the spectral index joining optical and X-ray emission,
$\beta=-1.05\pm 0.10$ (Halpern et al. 1999), is consistent with the
separate optical and X-ray 
indices.  From eq. (4) and the parameters for GRB 980519 given above, we
have  $\nu_c\approx 4\times 10^{16}$ Hz for $t_{\rm day}=1$, which is below
the X-ray frequency (3 keV = $7\times 10^{17}$ Hz).
We consider the agreement adequate in view of the expected gradual
turnover in flux and the observational and theoretical
uncertainty in the value of $\nu_c$.
The model does require a low value of $\epsilon_B$, as has also been
inferred in a number of other afterglows (Galama et al. 1999).

The model R-band flux densities on days 60 and 66 of Fig. 1 
fall  short of the observed values by nearly two orders
of magnitude; the emission 
on these two days is presumed to come from the host galaxy of
the burst (Sokolov et al. 1998; Bloom et al. 1998a). 
 However,  the 
observed flux densities on days 60 and 66 are  close 
to those expected of SN 1998bw at a cosmological distance of $z=1$
(see Fig. 1).
Despite excellent seeing 
conditions at Keck II, Bloom et al. (1998a) found little 
evidence for the extension 
expected of a host galaxy, so the presence of
 a SN 1998bw-type supernova is possible. 
Observations of the source at later times than have been reported should 
be able to confirm or reject this possibility.
 
Another burst with a relatively steep decline with time is GRB 980326,
in which the optical $F_{\nu}\propto t^{-2.10\pm 0.13}$ (Groot et al. 1998).
There is no radio data for this object, but the last optical observation
is a factor $\sim 10$ above the power law decline; this is not the host
galaxy because it is not present at a later time (Bloom \& Kulkarni 1998).
A possible source for the emission is a supernova (Bloom et al. 1999b),
as in the case of GRB 980519.
Such an event would be consistent with expansion into a wind and the
explosion of a massive star.
Bloom et al. (1999b) found that the spectrum of the late time source is
consistent with that of a supernova, which rules out the possibility of
a rise in the nonthermal afterglow emission as observed in
 GRB 970508  
(Panaitescu et al. 1998) and
the radio emission from SN 1998bw (Kulkarni et al. 1998; Li \& Chevalier 1999).

Based on the discovery of the SN 1998bw/GRB 980425 association,
Bloom et al. (1998b) proposed a subclass of GRBs produced by supernovae
(S-GRBs).
We also place SN 1998bw in a supernova class of
GRB, but with different properties
from those of Bloom et al. (1998b).
By supernova, we mean an observed event with an optical light curve
and spectrum similar to that of a Type I or Type II supernova.
In our picture, the GRB 990123 is an example of a GRB interacting with
the ISM (interstellar medium).
The similar (slow) rates of decline observed in other GRB afterglows suggests
that this type of object is the most common.
These objects are not
interacting with winds, do not have massive star progenitors, and are
not accompanied by supernovae.
Plausible progenitors are the mergers of compact objects.
The GRBs in the wind type are 
interacting with winds, have massive star progenitors, and are
 accompanied by supernovae.
SN 1998bw/GRB 980425 is the best example of this class of object.
GRB 980519 and GRB 980326 are other possible members.
Bloom et al. (1998b) propose that S-GRBs have no long-lived X-ray
afterglow because of synchrotron losses.
We suggest that X-ray afterglows are possible, although with steeper
time evolution than in the ISM case, provided $\epsilon_B$ is small,
as is also inferred for some of the ISM type afterglows.
Bloom et al. further propose that S-GRBs  have single pulse
GRB profiles.
We have found evidence for nonuniform energy input in SN 1998bw/GRB 980425
(Li \& Chevalier 1999) and suggest that both types of bursts
are powered by  central engines with irregular power output.
The GRB itself may not allow a classification of the event;  the two
cosmological GRBs mentioned as possible members of the wind class
have multiple peak time structure (in 't Zand et al. 1999; Groot et al. 1998).

\acknowledgments
We are grateful to J. Bloom, S. Kulkarni, and an anonymous referee
for useful comments and information.
Support for this work was provided in part by NASA grant NAG5-8232.

\clearpage

\begin{figure}

\caption{Wind interaction model for the afterglow
of GRB 980519. The radio, optical,
and X-ray data are taken, respectively, from Frail et al. (1998),
Halpern et al. (1999), and Nicastro et al. (1999). A factor-of-two
increase is applied to the original optical data (open squares) 
to account for dust extinction. The X-ray flux densities (large
filled squares) are the observed fluxes in the 2-10 kev band divided 
by the band width. The solid curves with small filled dots are the
light curves from a simple synchrotron self-absorption model 
described in the text. The open circles are the R-band flux densities
that SN 1998bw would have at a cosmological redshift of $z=1$ based
on its U-band flux densities at $z=0.0085$ (Galama et al. 1998). 
}

\end{figure}

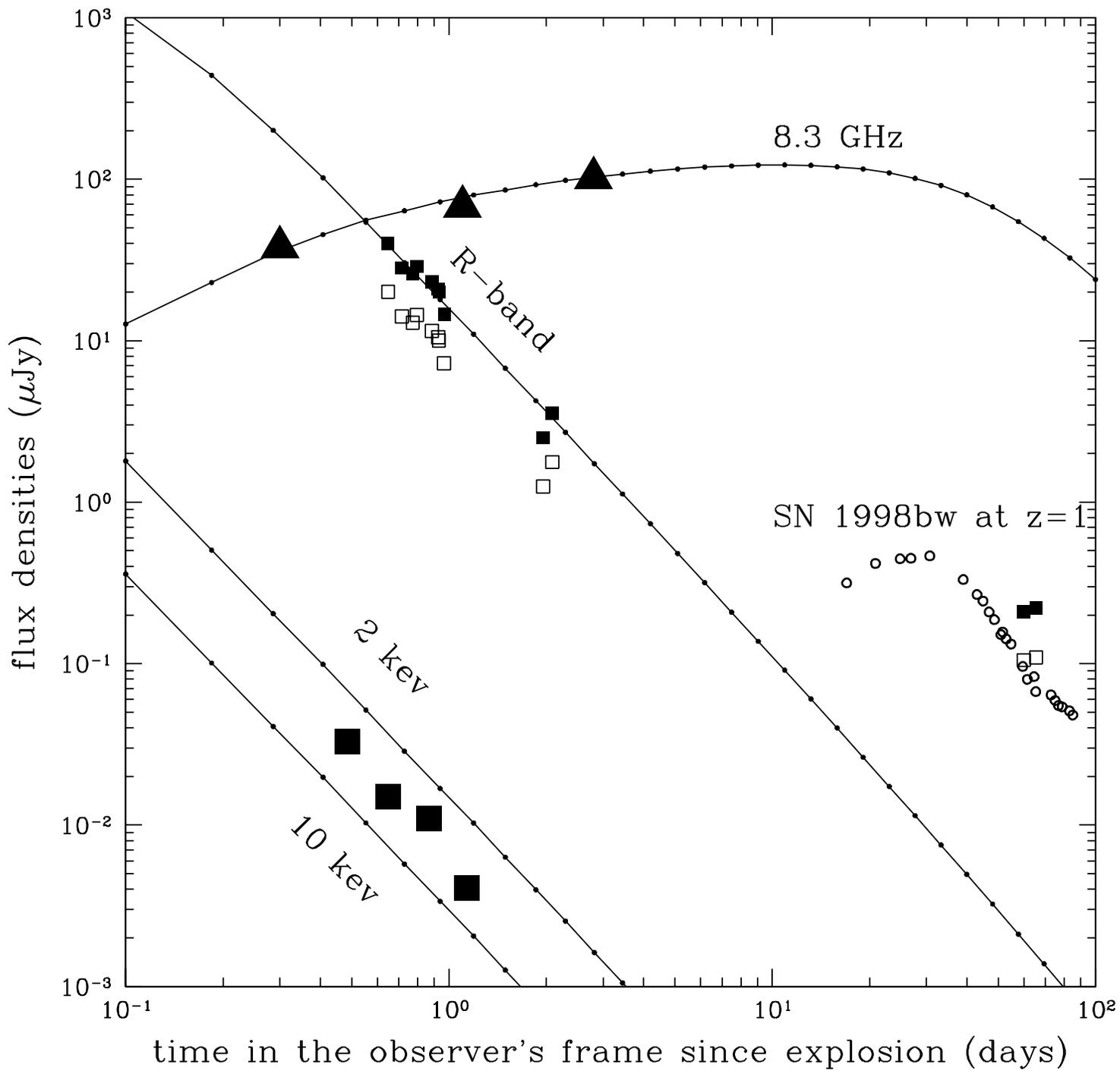
\end{document}